\documentclass[a4paper]{jpconf}

\usepackage{hyperref}

\usepackage{amsmath,amsfonts,amsthm,amssymb} % Math packages

\usepackage{tikz}

\usepackage{graphicx}
\graphicspath{ {images/} }
\usepackage{placeins}

\usepackage{algorithm}
\usepackage[noend]{algpseudocode}

\begin{document}
\title{Resampling schemes in population annealing -- numerical results}

\author{Denis Gessert$^{1,2}$, Martin Weigel$^{1,3}$ and Wolfhard Janke$^{2}$}

\address{$^1$ Centre for Fluid and Complex Systems, Coventry University, Coventry CV1 5FB, UK}
\address{$^2$ Institut f\"ur Theoretische Physik, Universität Leipzig, IPF 231101, 04081 Leipzig, Germany}
\address{$^3$ Institut f\"ur Physik, Technische Universität Chemnitz, 09107 Chemnitz, Germany}

\ead{gessertd@uni.coventry.ac.uk}

\begin{abstract}

Population annealing (PA) is a population-based algorithm that is designed for equilibrium simulations of thermodynamic systems with a rough free energy landscape. It is known to be more efficient in doing so than standard Markov chain Monte Carlo alone. The algorithm has a number of parameters that can be fine-tuned to improve performance. While there is some theoretical and numerical work regarding most of these parameters, there appears to be a gap in the literature concerning the role of resampling in PA. 
Here, we present a numerical comparison of a number of resampling schemes for PA simulations of the 2D Ising model.
\end{abstract}

\section{Introduction}
Systems with a rough free energy landscape, such as spin glasses, structural glasses and some polymers, to name a few, appear in many fields of research and are notoriously hard to simulate~\cite{Janke2008}. Over the last decades a number of methods have been developed with the aim of making such systems more accessible such as, e.g., multicanonical simulations~\cite{Berg1991}, the Wang-Landau algorithm~\cite{Wang2001}, parallel tempering~\cite{Hukushima1996} and the population annealing (PA) algorithm~\cite{Iba2001, Hukushima2003} which is the subject of the present work. 

While there is no clear-cut answer to the question which of these methods is best and the answer may depend on the system considered, we motivate the use of PA as follows: First, PA is well suited for the highly parallel execution on supercomputers, as the number of replicas is only limited by the underlying computing architecture and not the method itself. In contrast, in parallel tempering the level of parallelization is limited by the number of temperature points considered which is bounded by increasing round-trip times as the number of replicas grows. Additionally, with increasing population size not just the statistical error but also the bias is systematically reduced~\cite{Weigel2021}. Second, as part of the PA algorithm an equilibration routine is run at each temperature, which typically consists of a number of Markov chain Monte Carlo (MCMC) sweeps (MCS). However, the type of update is not specified and can freely be replaced by, e.g., a cluster algorithm, a rejection-free update or even by molecular dynamics~\cite{Christiansen2019a}. This allows for an easy adaptation of PA to the system under study, and any algorithmic improvement in MCMC immediately translates to PA.

The fundamental idea of PA, first presented by Hukushima and Iba~\cite{Iba2001,Hukushima2003}, is to perform parallel simulated annealing (SA) simulations (replicas) subject to a population-control step at each temperature. It is worth mentioning that thanks to this control step PA is an equilibrium method for calculating canonical averages although SA itself by default is not. 
The population control is implemented by a resampling phase employed at each temperature step (or in principle after $M$ steps, but researchers have focused on $M=1$) and is crucial to the performance of PA. Despite its importance to the success of the algorithm, it has received little attention in the literature -- the effect of different resampling methods has not yet been systematically studied and particularly the fact that resampling introduces additional artificial noise into the simulations has not been discussed.

We use the two-dimensional ferromagnetic Ising model in zero field corresponding to the Hamiltonian

\begin{equation}
    \mathcal{H} = -J \sum_{\langle ij \rangle} \sigma_i \sigma_j
\end{equation}
where the sum is over nearest neighbors $\langle ij \rangle$ only. In our simulations we consider lattices of size $64 \times 64$ with periodic boundary conditions. Our implementation of the PA algorithm is based on the Open Source GPU code of Ref.~\cite{Barash2017} that is available online. The code uses a checkerboard decomposed Metropolis update as an equilibration routine. All resampling methods were also implemented on GPUs.

\section{Simulation method}
\label{sec:methods}
\subsection{Population annealing}
\label{sec:PA}
\begin{figure}%[ht]
    \centering
    \vspace{-0.5cm}
    \includegraphics[width=0.8499\textwidth]{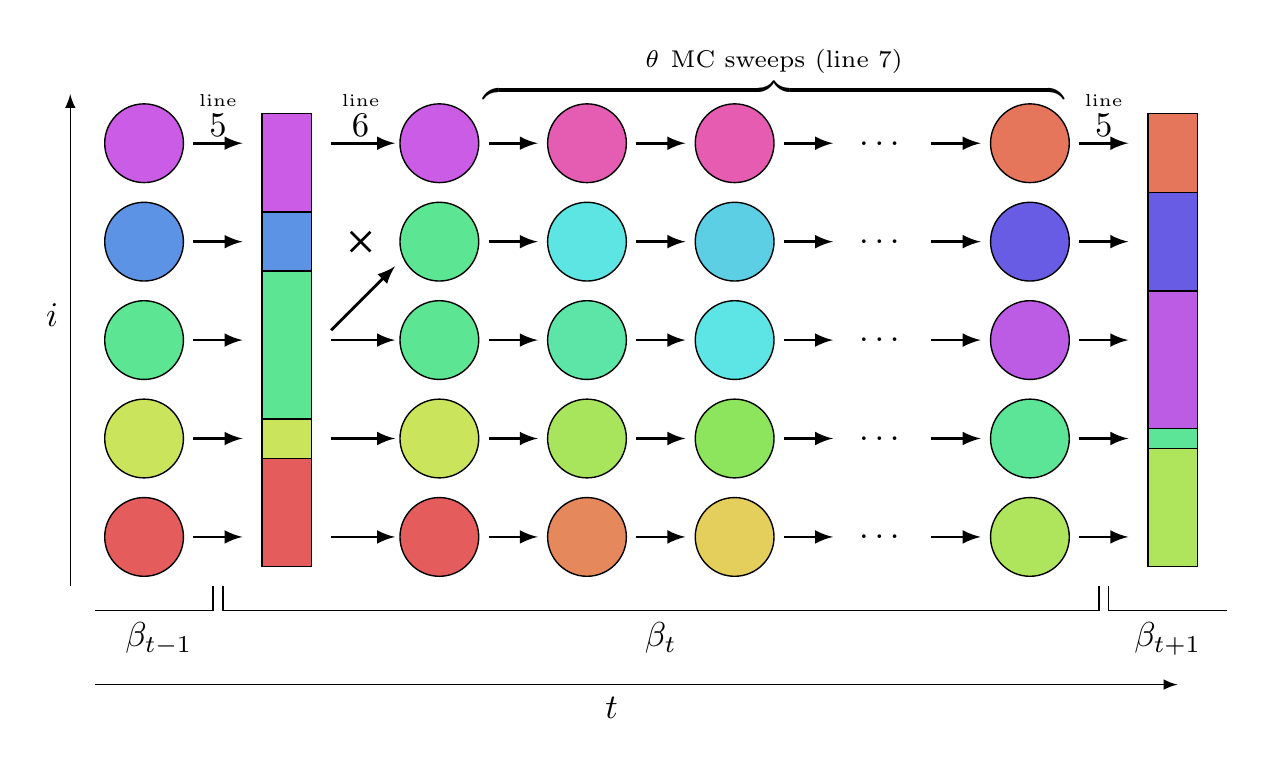}
    \vspace{-0.75cm}
    \caption{\label{fig:PAscheme}Schematic of the population annealing algorithm. Vertical axis corresponds to $R$ replicas (here five) and horizontal axis to simulation time. Line numbers correspond to algorithm~\ref{alg:PA} in \ref{sec:PAAlgo}.}
\end{figure}

In PA a population of $R_0=R$ replicas is initialized at the starting inverse temperature $\beta_0$, chosen to be zero if possible; here, $R$ is the target population size.
What follows is an iteration scheme (see Figure~\ref{fig:PAscheme}) in which in each iteration the inverse temperature~$\beta_t$ is increased to $\beta_{t+1}=\beta_t + \Delta \beta$. The step $\Delta \beta$ has to be chosen appropriately and can in principle be adaptive~\cite{Barash2017,Christiansen2019}.

At the end of iteration $t-1$, the population consists of $R_{t-1}$ replicas at $\beta_{t-1}$. Following the line of thought of histogram reweighting~\cite{Ferrenberg1988}, if we were asked to estimate the energy histogram at $\beta_t$ based on the population at $\beta_{t-1}$, we would attribute a weight $w_i \propto \exp(- \Delta\beta E_i)$ to each replica $i$ with energy $E_i$ where the weights are normalized such that they sum to one.
Next, the population is resampled according to these weights $w_i$, while attempting to keep the population size close to the target size $R$. In practice, the number of descendants $r_i$ of replica $i$ is, on average,  
\begin{equation}
    \langle r_i \rangle = \hat \tau_i = w_i R. \label{equ:TauHatDef}
\end{equation}
Clearly, this only specifies the mean of $r_i$ and the exact way how $r_i$ is determined depends on the used resampling method. Note that it is common that $r_i = 0$ for some replicas, in which case they are culled while others are replicated multiple times.

Immediately after resampling, the population is potentially highly correlated as it may contain identical configurations. To reduce the correlation among replicas, an equilibration routine is used (here a checkerboard decomposed version of the Metropolis algorithm, see Ref.~\cite{Barash2017} for details). After $\theta$ MCS, population averages for observables of interest are calculated.
The scheme then proceeds with the iteration for inverse temperature $\beta_{t+1}$ until the final temperature is reached.

For a pseudocode implementation of the PA algorithm we refer to \ref{sec:PAAlgo}.

\subsection{Resampling methods}
\label{sec:resamplingMethods}
\begin{figure}[b]
    \centering
    % \begin{minipage}{0.48\textwidth}
    % \hfil\hspace{0.1\textwidth}\hfil
    \begin{minipage}{0.38\textwidth}
        \centering
    \resizebox{\textwidth}{!}{    \begin{tikzpicture}
    
        \draw[fill=teal!30,draw=teal!30] (-4,0) rectangle (-2.9269120986876436,1); 
        \draw[fill=red!30,draw=red!30] (-2.9269120986876436,0) rectangle (-2.4822674553394175,1); 
        \draw[fill=yellow!30,draw=yellow!30] (-2.4822674553394175,0) rectangle (-0.9446526792490118,1); 
        \draw[fill=blue!30,draw=blue!30] (-0.9446526792490118,0) rectangle (-0.6951478498549886,1); 
        \draw[fill=green!30,draw=green!30] (-0.6951478498549886,0) rectangle (-0.04204881460788279,1); 
        \draw[fill=gray!30,draw=gray!30] (-0.04204881460788279,0) rectangle (1.3962030604563669,1); 
        \draw[fill=orange!30,draw=orange!30] (1.3962030604563669,0) rectangle (2.3749195701022012,1); 
        \draw[fill=magenta!30,draw=magenta!30] (2.3749195701022012,0) rectangle (3.810497626545277,1); 
        \draw[fill=brown!30,draw=brown!30] (3.810497626545277,0) rectangle (4.590802717990806,1); 
        \draw[fill=lime!30,draw=lime!30] (4.590802717990806,0) rectangle (6,1); 
    
        \draw[step=1] (-4,0) grid (6,1);
    
        \draw[very thin,draw=lightgray] (-4,-0.85) -- (-4,0);
        \draw[very thin,draw=lightgray] (6,-0.85) -- (6,0);
        \draw[latex-latex] (-4,-0.75) -- (1,-0.75) node[below]{$R=10$} -- (6,-0.75);
    
        \draw[-latex,thick] (-3.211844615390551,1.5) node [above] {$U_1$} -- (-3.211844615390551,0.9);
        \draw[-latex,thick] (5.263803059554949,1.5) node [above] {$U_2$} -- (5.263803059554949,0.9);
        \draw[-latex,thick] (1.7618665388539627,1.5) node [above] {$U_3$} -- (1.7618665388539627,0.9);
        \draw[-latex,thick] (-2.5102698594643114,1.5) node [above] {} -- (-2.5102698594643114,0.9);
        \draw[-latex,thick] (5.631515266160363,1.5) node [above] {} -- (5.631515266160363,0.9);
        \draw[-latex,thick] (5.952117527909541,1.5) node [above] {} -- (5.952117527909541,0.9);
        \draw[-latex,thick] (0.7306485892515528,1.5) node [above] {} -- (0.7306485892515528,0.9);
        \draw[-latex,thick] (-1.2768833224926288,1.5) node [above] {$U_8$} -- (-1.2768833224926288,0.9);
        \draw[-latex,thick] (-3.9249844929708164,1.5) node [above] {$U_9$} -- (-3.9249844929708164,0.9);    
        \draw[-latex,thick] (0.5690815377134548,1.5) node [above] {$U_{10}$} -- (0.5690815377134548,0.9);
    \end{tikzpicture}} \\
    (a) multinomial
    \resizebox{\textwidth}{!}{    \begin{tikzpicture}
    
        \draw[fill=teal!30,draw=teal!30] (-4,0) rectangle (-2.9269120986876436,1); 
        \draw[fill=red!30,draw=red!30] (-2.9269120986876436,0) rectangle (-2.4822674553394175,1); 
        \draw[fill=yellow!30,draw=yellow!30] (-2.4822674553394175,0) rectangle (-0.9446526792490118,1); 
        \draw[fill=blue!30,draw=blue!30] (-0.9446526792490118,0) rectangle (-0.6951478498549886,1); 
        \draw[fill=green!30,draw=green!30] (-0.6951478498549886,0) rectangle (-0.04204881460788279,1); 
        \draw[fill=gray!30,draw=gray!30] (-0.04204881460788279,0) rectangle (1.3962030604563669,1); 
        \draw[fill=orange!30,draw=orange!30] (1.3962030604563669,0) rectangle (2.3749195701022012,1); 
        \draw[fill=magenta!30,draw=magenta!30] (2.3749195701022012,0) rectangle (3.810497626545277,1); 
        \draw[fill=brown!30,draw=brown!30] (3.810497626545277,0) rectangle (4.590802717990806,1); 
        \draw[fill=lime!30,draw=lime!30] (4.590802717990806,0) rectangle (6,1); 
    
        \draw[step=1] (-4,0) grid (6,1);
    
        \draw[very thin,draw=lightgray] (-4,-0.85) -- (-4,0);
        \draw[very thin,draw=lightgray] (6,-0.85) -- (6,0);
        \draw[latex-latex] (-4,-0.75) -- (1,-0.75) node[below]{$R=10$} -- (6,-0.75);
    
        \draw[-latex,thick] (-3.7400317686381097,1.5) node [above] {$U_{1}$} -- (-3.7400317686381097,0.9);
        \draw[-latex,thick] (-2.0222055879968535,1.5) node [above] {$U_{2}$} -- (-2.0222055879968535,0.9);
        \draw[-latex,thick] (-1.6511484153148008,1.5) node [above] {$U_{3}$} -- (-1.6511484153148008,0.9);
        \draw[-latex,thick] (-0.5620589560480992,1.5) node [above] {} -- (-0.5620589560480992,0.9);
        \draw[-latex,thick] (0.8797598007899547,1.5) node [above] {} -- (0.8797598007899547,0.9);
        \draw[-latex,thick] (1.7734380043055271,1.5) node [above] {} -- (1.7734380043055271,0.9);
        \draw[-latex,thick] (2.5398901870625465,1.5) node [above] {} -- (2.5398901870625465,0.9);
        \draw[-latex,thick] (3.0220435666939873,1.5) node [above] {$U_{8}$} -- (3.0220435666939873,0.9);
        \draw[-latex,thick] (4.384682607272419,1.5) node [above] {$U_{9}$} -- (4.384682607272419,0.9);
        \draw[-latex,thick] (5.749267824203895,1.5) node [above] {$U_{10}$} -- (5.749267824203895,0.9);
    \end{tikzpicture}} \\
    (c) stratified
    \end{minipage}\hfil\hspace{0.02\textwidth}\hfil
    \begin{minipage}{0.4\textwidth}
        \centering
        \resizebox{\textwidth}{!}{    \begin{tikzpicture}
    
        \draw[fill=teal!30,draw=teal!30] (-4,0) rectangle (-2.9269120986876436,1); 
        \draw[fill=red!30,draw=red!30] (-2.9269120986876436,0) rectangle (-2.4822674553394175,1); 
        \draw[fill=yellow!30,draw=yellow!30] (-2.4822674553394175,0) rectangle (-0.9446526792490118,1); 
        \draw[fill=blue!30,draw=blue!30] (-0.9446526792490118,0) rectangle (-0.6951478498549886,1); 
        \draw[fill=green!30,draw=green!30] (-0.6951478498549886,0) rectangle (-0.04204881460788279,1); 
        \draw[fill=gray!30,draw=gray!30] (-0.04204881460788279,0) rectangle (1.3962030604563669,1); 
        \draw[fill=orange!30,draw=orange!30] (1.3962030604563669,0) rectangle (2.3749195701022012,1); 
        \draw[fill=magenta!30,draw=magenta!30] (2.3749195701022012,0) rectangle (3.810497626545277,1); 
        \draw[fill=brown!30,draw=brown!30] (3.810497626545277,0) rectangle (4.590802717990806,1); 
        \draw[fill=lime!30,draw=lime!30] (4.590802717990806,0) rectangle (6,1); 
    
        \draw[step=1] (-4,0) grid (6,1);
    
        \draw[very thin,draw=lightgray] (-4,-0.85) -- (-4,0);
        \draw[very thin,draw=lightgray] (6,-0.85) -- (6,0);
        \draw[latex-latex] (-4,-0.75) -- (1,-0.75) node[below]{$R=10$} -- (6,-0.75);
    
        \draw[-latex,thick] (-3.8,1.5) node [above] {$U_1$} -- (-3.8,0.9);
        \draw[-latex,thick] (-2.8,1.5) node [above] {$U_2$} -- (-2.8,0.9);
        \draw[-latex,thick] (-1.8,1.5) node [above] {$U_3$} -- (-1.8,0.9);
        \draw[-latex,thick] (-0.8,1.5) node [above] {} -- (-0.8,0.9);
        \draw[-latex,thick] (0.2,1.5) node [above] {} -- (0.2,0.9);
        \draw[-latex,thick] (1.2,1.5) node [above] {} -- (1.2,0.9);
        \draw[-latex,thick] (2.2,1.5) node [above] {} -- (2.2,0.9);
        \draw[-latex,thick] (3.2,1.5) node [above] {$U_8$} -- (3.2,0.9);
        \draw[-latex,thick] (4.2,1.5) node [above] {$U_9$} -- (4.2,0.9);
        \draw[-latex,thick] (5.2,1.5) node [above] {$U_{10}$} -- (5.2,0.9);
    \end{tikzpicture}} \\
        (b) systematic \\
        \vspace{0.2cm}
        \resizebox{\textwidth}{!}{    \begin{tikzpicture}
    
        \draw[fill=teal!30,draw=teal!30] (-4,0) rectangle (-2.9269120986876436,1); 
        \draw[fill=red!30,draw=red!30] (-2.9269120986876436,0) rectangle (-2.4822674553394175,1); 
        \draw[fill=yellow!30,draw=yellow!30] (-2.4822674553394175,0) rectangle (-0.9446526792490118,1); 
        \draw[fill=blue!30,draw=blue!30] (-0.9446526792490118,0) rectangle (-0.6951478498549886,1); 
        \draw[fill=green!30,draw=green!30] (-0.6951478498549886,0) rectangle (-0.04204881460788279,1); 
        \draw[fill=gray!30,draw=gray!30] (-0.04204881460788279,0) rectangle (1.3962030604563669,1); 
        \draw[fill=orange!30,draw=orange!30] (1.3962030604563669,0) rectangle (2.3749195701022012,1); 
        \draw[fill=magenta!30,draw=magenta!30] (2.3749195701022012,0) rectangle (3.810497626545277,1); 
        \draw[fill=brown!30,draw=brown!30] (3.810497626545277,0) rectangle (4.590802717990806,1); 
        \draw[fill=lime!30,draw=lime!30] (4.590802717990806,0) rectangle (6,1); 
    
        \draw[step=1] (-4,0) grid (6,1);
    
        \draw[very thin,draw=lightgray] (-4,-0.85) -- (-4,0);
        \draw[very thin,draw=lightgray] (6,-0.85) -- (6,0);
        \draw[latex-latex] (-4,-0.75) -- (1,-0.75) node[below]{$R=10$} -- (6,-0.75);

        \draw[fill=teal!30,draw=teal!30] (2,-3) rectangle (6-3.9269120986876436,-4); 
        \draw[fill=red!30,draw=red!30] (6-3.9269120986876436,-3) rectangle (6-3.4822674553394175,-4); 
        \draw[fill=yellow!30,draw=yellow!30] (6-3.4822674553394175,-3) rectangle (6-2.9446526792490118,-4); 
        \draw[fill=blue!30,draw=blue!30] (6-2.9446526792490118,-3) rectangle (6-2.6951478498549886,-4); 
        \draw[fill=green!30,draw=green!30] (6-2.6951478498549886,-3) rectangle (6-2.042048814607883,-4); 
        \draw[fill=gray!30,draw=gray!30] (6-2.042048814607883,-3) rectangle (6-1.6037969395436331,-4); 
        \draw[fill=orange!30,draw=orange!30] (6-1.6037969395436331,-3) rectangle (6-0.6250804298977988,-4); 
        \draw[fill=magenta!30,draw=magenta!30] (6-0.6250804298977988,-3) rectangle (6-0.1895023734547232,-4); 
        \draw[fill=brown!30,draw=brown!30] (6-0.1895023734547232,-3) rectangle (6.5908027179908064,-4); 
        \draw[fill=lime!30,draw=lime!30] (6.5908027179908064,-3) rectangle (7,-4); 
    
        \draw[fill=teal!30,draw=teal!30] (-4,-3) rectangle (-3,-4); 
        \draw[fill=yellow!30,draw=yellow!30] (-3,-3) rectangle (-2,-4); 
        \draw[fill=gray!30,draw=gray!30] (-2,-3) rectangle (-1,-4); 
        \draw[fill=magenta!30,draw=magenta!30] (-1,-3) rectangle (0,-4); 
        \draw[fill=lime!30,draw=lime!30] (0,-3) rectangle (1,-4); 
    
        \draw[step=1] (-4,-3) grid (1,-4);
        \draw[step=1] (2,-3) grid (7,-4);
        
        \draw[line width=2mm,-latex,red!20] (0,-0.5) -- (-2,-2.5);
        \draw[line width=2mm,-latex,red!20] (2,-0.5) -- (4,-2.5);

        \draw[-latex,thick] (5.99354058740457,-2.5) node [above] {$U_1$} -- (5.99354058740457,-3.1);
        \draw[-latex,thick] (6.410456264398866,-2.5) node [above] {$U_2$} -- (6.410456264398866,-3.1);
        \draw[-latex,thick] (3.0240457800207676,-2.5) node [above] {$U_3$} -- (3.0240457800207676,-3.1);
        % \draw[-latex,thick] (2.607401748642331,-2.5) node [above] {$U_4$} -- (2.607401748642331,-3.1);
        \draw[-latex,thick] (5.607401748642331,-2.5) node [above] {$U_4$} -- (5.607401748642331,-3.1);
        \draw[-latex,thick] (2.272019702643067,-2.5) node [above] {$U_5$} -- (2.272019702643067,-3.1);
    \end{tikzpicture}} \\
        (d) residual
    \end{minipage}
    \caption{Visualization of various population size preserving resampling methods. Colored boxes correspond to different replicas $i$ and their lengths are proportional to $\hat \tau_i$. The number of copies of replica~$i$ is determined by the number of arrows in box $i$. (a) - (c) $R$ arrows are distributed over the interval $[0,R]$. (a) multinomial: Each arrow is uniformly distributed over the interval $[0,R]$. (b) systematic: The first arrow is uniformly distributed within $[0,1]$. The remaining $R-1$ arrows are placed with unit spacing. (c) stratified: In each square exactly one arrow is placed with uniform probability, i.e., $U_i \sim \mathcal U ([i-1,i])$. (d) residual: At first each replica is copied $\lfloor\hat\tau_i\rfloor$ times. The population is brought to its original size by multinomially drawing from the residuals, i.e., performing (a) with $\tau_i - \lfloor\tau_i\rfloor$ replacing $\tau_i$.}
    \label{fig:VisResamplingPreservedPopulation}
\end{figure}
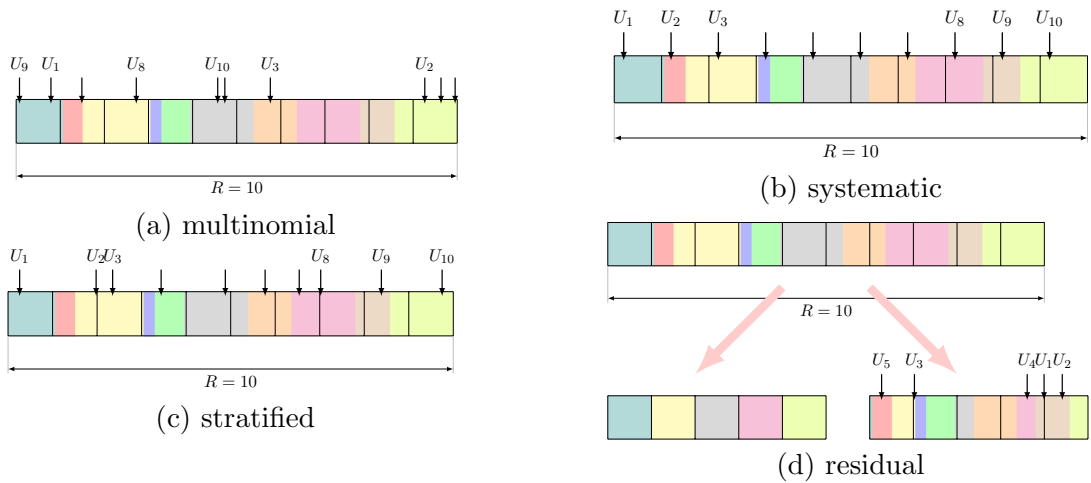
    
When the number of copies of one replica $r_i$ is chosen only based on the value of $\hat\tau_i$ of the same replica~$i$, the resampling method is completely described by a univariate distribution $P_{\hat\tau_i}(r_i = k)$. In this case the sum of all $r_i$'s is a random variable and thus the population size fluctuates with time. In this group we consider ``nearest integer'' resampling~\cite{Wang2015} and Poisson resampling~\cite{Machta2010} given by 
\begin{equation}
    P_{\hat\tau_i}(r_i = k) = \begin{cases}
        \hat\tau_i - \lceil\hat\tau_i \rceil & \text{if } k = \lceil\hat\tau_i \rceil \\
        1 - (\hat\tau_i - \lceil\hat\tau_i \rceil) & \text{if } k = \lfloor\hat\tau_i \rfloor \\
        0 & \text{else}
    \end{cases}
\end{equation}
for nearest-integer and
\begin{equation}
    P_{\hat\tau_i}(r_i = k) = \frac{\hat\tau_i^{k}}{k!} \, e^{-\hat\tau_i }\hspace{2.1cm}
\end{equation}
for Poisson, respectively. Here, $\lceil x \rceil$ denotes the smallest integer that is larger than or equal to $x$, i.e., rounding up, and similarly $\lfloor x \rfloor$ denotes the largest integer smaller than or equal to $x$, i.e., rounding down.
    
Vice versa, if one requires the population size to be fixed, the set $\{r_i\}$ is drawn at once and the individual $r_i$'s may show some correlation. See Figure~\ref{fig:VisResamplingPreservedPopulation} for a visualization of the population size preserving resampling methods used here and the caption for explanations. These methods are quite well known in the field of particle filters~\cite{doucet2009tutorial} whereas within the PA community multinomial resampling is the only population size preserving method used so far.

In principle, one might come up with many more methods such as geometric resampling. Here we restrict ourselves to methods with sampling variance less than or equal to that of the widely used multinomial resampling.

\subsection{Benchmarking quantities}
\label{sec:benchmarkingQuantities}
The most obvious quantities to use for comparing the algorithmic performance are systematic and statistical errors. We looked at various observables and present here the errors measured for the specific heat $C_v$, as this is where we found the strongest differences among the methods.

Additionally, Ref.~\cite{Wang2015} introduced population quantities designed to identify the maximal correlation within the population introduced through the resampling step, i.e.,

\begin{equation}
    \rho_t = R \sum_{i=1}^{R} \mathfrak{n}_i^2, \quad \quad \rho_s = R \exp\left( \sum_{i=1}^R \mathfrak{n}_i \ln \mathfrak{n}_i\right) \quad \text{ and } \quad f = \sum_{i=1} \min\{1,R\mathfrak{n}_i \},
\end{equation}

\noindent where $\mathfrak{n}_i$ is the fraction of replicas descending from the initial replica $i$, $\rho_t$ is known as the mean square family size, $\rho_s$ as the entropic family size and $f$ is the number of families. As the authors of Ref.~\cite{Wang2015} show, these quantities are closely related to each other.
\section{Results}
\label{sec:results}

The following results were obtained from simulations of the Ising model on a square lattice with periodic boundary conditions of dimensions $64\times 64$. The numerical data presented here was attained through independent PA simulations for each resampling method and was averaged over $5\,000$ simulations per method. The target population size is $R=20\,000$ and the annealing schedule was chosen to be $\beta_k = k / 300$ with $k\in\{0,\dots,299\}$. The choice of $\beta_0=0$ allows us to initialize each spin configuration simply by choosing every spin as up or down with equal probabilities which can easily be implemented. At each temperature $\theta=5$ MCS were carried out.

\begin{figure}[t]    
    \centering
    {\hspace{-0.5cm} \includegraphics{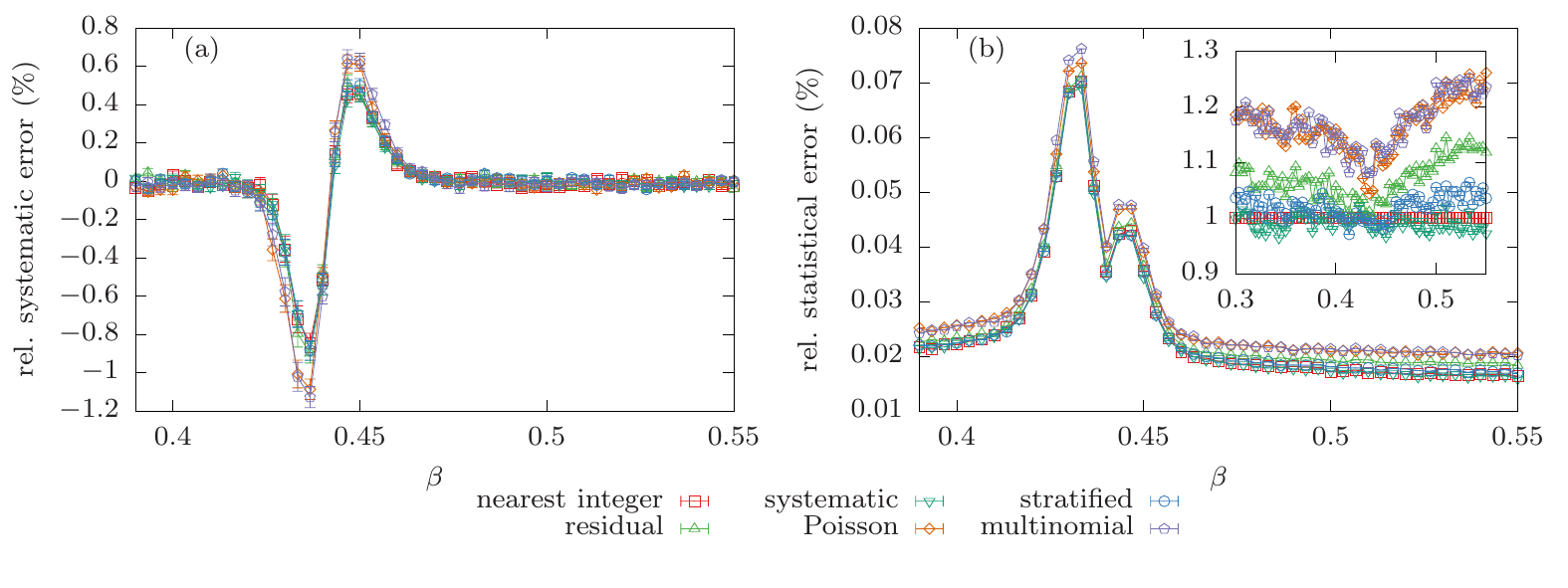}}
    \vspace{-0.8cm}
    \caption{Relative (a) systematic and (b) statistical error of the specific heat measured after the resampling step for various resampling methods. (a)  Systematic errors hardly differ between different resampling schemes. (b) The inset shows the statistical error relative to the error using nearest-integer resampling. Away from criticality, the curves differ significantly whereas around $\beta_c\approx 0.44$ the choice of the resampling method appears to have little effect on the statistical error.}
    \vspace{-0.21cm}
    \label{fig:SW_specHeat}
\end{figure}

Figure~\ref{fig:SW_specHeat} shows bias and statistical error immediately after resampling (before the equilibration routine) of the specific heat. We calculate the bias by using the known exact solution~\cite{Beale1996, Kaufmann1949}, whereas the statistical errors were obtained by estimating the standard deviation of independent runs. Both quantities peak around the critical temperature. The bias shows little deviation among the methods, although multinomial and Poisson resampling appear to have slightly stronger bias in the vicinity of the critical temperature. Around the critical temperature statistical errors are almost indistinguishable and away from criticality the statistical error using multinomial and Poisson resampling is significantly larger compared to the other methods. The inset, which shows all errors normalized to the nearest-integer errors, illustrates that Poisson and multinomial resampling have statistical errors of up to 25\% higher than nearest-integer resampling and outside the depicted near-critical regime these even exceed 30\%. We argue that near criticality fluctuations intrinsic to the model are very strong and dominate the observed statistical errors whereas further away the resampling noise has a significant effect.

\begin{figure}[b]
    \centering
    \includegraphics{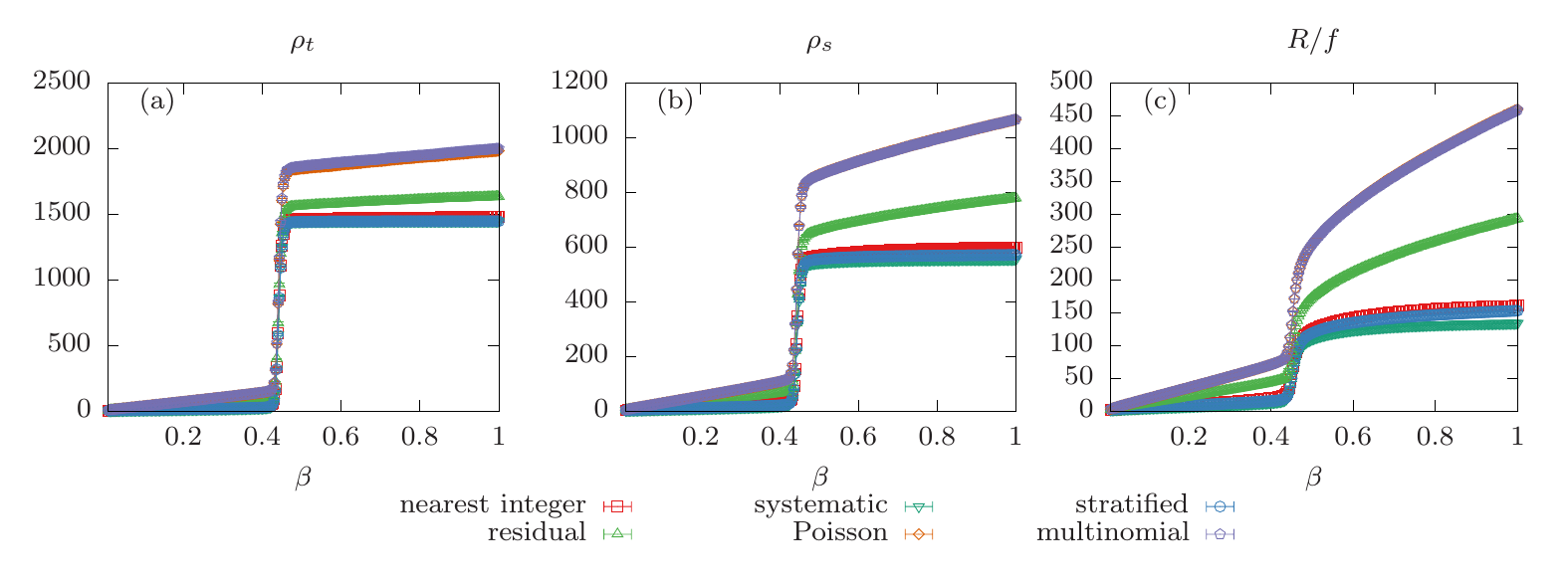}
    \vspace{-1cm}
    \caption{Family quantities (a) $\rho_t$, (b) $\rho_s$ and (c) $R/f$ for various resampling methods.}% For readability sake two out of three data points were omitted.}
    \label{fig:familyQu}
\end{figure}

The family quantity $\rho_t$ is shown in Figure~\ref{fig:familyQu}(a). It increases rapidly around the critical point and slowly otherwise. It agrees with the previous observations in the sense that again multinomial and Poisson resampling stick out and $\rho_t$ consistently takes its highest values for these two methods. Larger average family size typically corresponds to more correlation within the population and thus poorer statistics. The same observations are made for the entropic family size $\rho_s$ and the inverse of the number of families $R/f$.

\section{Conclusion}
\label{sec:conclusion}
We have demonstrated that resampling has a significant effect on the data obtained through population annealing, and we see a difference in error bars of up to 30\% using standard resampling methods. Furthermore, we identified that resampling matters most away from criticality when it contributes as much or more to the statistical error as the fluctuations intrinsic to the system. Through the study of family quantities we found that higher variance in $r_i$ at high temperatures causes fewer families of larger sizes surviving, thus causing more correlation in the population.

In a quest of finding the best resampling method nearest-integer resampling is the front runner due to it consistently outperforming the other methods and due to its simple implementation. Systematic resampling is also a good choice, in particular if a constant population size is desired, as it provides similar performance although its implementation is slightly more involved. In particular, we note that it performs significantly better than the multinomial method usually used for this purpose.

\ack
The project was supported by the Deutsch-Französische Hochschule (DFH-UFA) through the Doctoral College ``$\mathbb{L}^4$'' under Grant No. CDFA-02-07.

\appendix
\section{Population Annealing algorithm}
\label{sec:PAAlgo}
\begin{algorithm}[H]
    \caption{Standard Population Annealing algorithm (introduced in Ref.~\cite{Hukushima2003})}
    \label{alg:PA}
    \begin{algorithmic}[1]
      \Require $R$,$\{\beta_k\}$,$\theta$, $M$
    %   \Procedure {PopAnnealing}{$R$,$\{\beta_k\}$,$\theta$, $M$}
      \State $k \leftarrow 0$\;
      \State Initialize population of $R$ independent replicas at $\beta_0$\;
      \While{$\beta_{k+1} < \beta_{\max}$}
        \State $k \leftarrow t+1$\;
        \State Calculate \textbf{Boltzmann weights} $w_i$ of the replicas at $\beta_t$\;
        \State \textbf{if} $k \equiv 0( \textrm{mod } M)$ then \textbf{resample} population according to weights 
        
        \hspace{4cm} (on avg. $\tau_i=R w_i$ copies of replica $i$)\;
        \State \textbf{Monte Carlo update} of the replicas ($\theta$ MC sweeps at $\beta_k$)\;
        \State \textbf{Measure} observables $\mathcal O$\;
      \EndWhile
    %   \EndProcedure
    \end{algorithmic}
\end{algorithm}
Algorithm~\ref{alg:PA} shows the standard population annealing algorithm as originally devised by Hukushima and Iba~\cite{Hukushima2003}. The parameters are the target population size $R$ (throughout the simulation $R_i$ might fluctuate around $R$), the annealing schedule $\{\beta_k\}$, the number of equilibration  updates $\theta$ (typically MCMC) between each temperature step and $M$, the resampling interval (usually $M=1$).

\section*{References}
\bibliography{references}

\providecommand{\newblock}{}
\begin{thebibliography}{10}
\expandafter\ifx\csname url\endcsname\relax
  \def\url#1{{\tt #1}}\fi
\expandafter\ifx\csname urlprefix\endcsname\relax\def\urlprefix{URL }\fi
\providecommand{\eprint}[2][]{\url{#2}}

\bibitem{Janke2008}
Janke W (ed) 2007 {\em Rugged Free Energy Landscapes -- Common Computational
  Approaches to Spin Glasses, Structural Glasses and Biological
  Macromolecules\/} ({\em Lecture Notes in Physics\/} vol 736) (Berlin:
  Springer)

\bibitem{Berg1991}
Berg B~A and Neuhaus T 1991 {\em Phys. Lett. B\/} {\bf 267} 249--53

\bibitem{Wang2001}
Wang F and Landau D~P 2001 {\em Phys. Rev. Lett.\/} {\bf 86} 2050--3

\bibitem{Hukushima1996}
Hukushima K and Nemoto K 1996 {\em J. Phys. Soc. Jpn.\/} {\bf 65} 1604--8

\bibitem{Iba2001}
Iba Y 2001 {\em Trans. Jpn. Soc. Artif. Intell.\/} {\bf 16} 279--86

\bibitem{Hukushima2003}
Hukushima K and Iba Y 2003 {\em {AIP} Conf. Proc.\/} {\bf 690} 200--6

\bibitem{Weigel2021}
Weigel M, Barash L, Shchur L and Janke W 2021 {\em Phys. Rev. E\/} {\bf 103}
  053301

\bibitem{Christiansen2019a}
Christiansen H, Weigel M and Janke W 2019 {\em Phys. Rev. Lett.\/} {\bf 122}
  060602

\bibitem{Barash2017}
Barash L~Y, Weigel M, Borovsk{\'{y}} M, Janke W and Shchur L~N 2017 {\em
  Comput. Phys. Commun.\/} {\bf 220} 341--50

\bibitem{Christiansen2019}
Christiansen H, Weigel M and Janke W 2019 {\em J. Phys.: Conf. Ser.\/} {\bf
  1163} 012074

\bibitem{Ferrenberg1988}
Ferrenberg A~M and Swendsen R~H 1988 {\em Phys. Rev. Lett.\/} {\bf 61} 2635--8

\bibitem{Wang2015}
Wang W, Machta J and Katzgraber H~G 2015 {\em Phys. Rev. E\/} {\bf 92} 063307

\bibitem{Machta2010}
Machta J 2010 {\em Phys. Rev. E\/} {\bf 82} 026704

\bibitem{doucet2009tutorial}
Doucet A and Johansen A~M 2011 {\em Handbook of Nonlinear Filtering\/} Crisan D
  and Rozovskii B (eds) (New York: Oxford University Press) chap 8.2, pp
  656--704

\bibitem{Beale1996}
Beale P~D 1996 {\em Phys. Rev. Lett.\/} {\bf 76} 78--81

\bibitem{Kaufmann1949}
Kaufman B 1949 {\em Phys. Rev.\/} {\bf 76} 1232--43

\end{thebibliography}

\end{document}